\documentclass[epj]{svjour}
\usepackage{graphicx}
\usepackage{amssymb}
\usepackage[square,comma,sort&compress]{natbib}
\begin{document}
\title{
Search for \boldmath $\Theta(1540)^+$ in exclusive proton-induced reaction
$p+\mathrm{C}(N)\to \Theta^+ \bar{K}^0 + \mathrm{C}(N)$ at the energy of 70 GeV
}
\subtitle{The SPHINX Collaboration 
}
\author{
Yu.M.~Antipov\inst{1} \and
A.V.~Artamonov\inst{1} \and
V.A.~Batarin\inst{1} \and
O.V.~Eroshin\inst{1} \and
V.Z.~Kolganov\inst{2} \and
A.S.~Konstantinov\inst{1} \and
A.P.~Kozhevnikov\inst{1} \and
V.F.~Kurshetsov\inst{1} \and
A.E.~Kushnirenko\inst{1} \and
L.G.~Landsberg\inst{1} \and
V.M.~Leontiev\inst{1} \and
G.S.~Lomkatsi\inst{2} \and
V.A.~Mukhin\inst{1} \and
V.V.~Molchanov\inst{1} \and
A.P.~Nilov\inst{2} \and
D.I.~Patalakha\inst{1} \and
S.V.~Petrenko\inst{1} \and
A.I.~Petrukhin\inst{1} \and
V.T.~Smolyankin\inst{2} \and
D.V.~Vavilov\inst{1} \and
V.A.~Victorov\inst{1} 
}                     
\institute{Institute for High Energy Physics, Protvino, Russia \and
Institute of Theoretical and Experimental Physics, Moscow, Russia}
\authorrunning{Yu.M.~Antipov et al.}
\titlerunning{Search for $\Theta^+(1540)$}
%
%
\abstract{
A search for narrow $\Theta(1540)^+$, a candidate for pentaquark
baryon with positive strangeness, has been performed in an exclusive 
proton-induced reaction
$p+\mathrm{C}(N)\to \Theta^+ \bar{K}^0 + \mathrm{C}(N)$
on carbon nuclei or quasifree nucleons at
$E_{\mathrm{beam}}$=~70~GeV ($\sqrt{s} = 11.5$~GeV) 
studying  $nK^+$, $pK_S^0$ and $pK_L^0$  decay channels
of $\Theta(1540)^+$
in four different final states of the $\Theta^+ \bar{K}^0$ system.
In order to assess the quality of the identification of the final states
with neutron or $K^0_L$ we reconstructed $\Lambda(1520)\to nK^0_S$ and
$\phi\to K^0_LK^0_S$ decays in the calibration reactions 
$p+\mathrm{C}(N)\to \Lambda(1520)K^+ + \mathrm{C}(N)$ and
$p+\mathrm{C}(N)\to p\phi + \mathrm{C}(N)$.
We found no evidence for narrow pentaquark peak in any of the studied final
states and decay channels.
Assuming that the production characteristics of the $\Theta^+ \bar{K^0}$ system
are not drastically different from those of the $\Lambda(1520)K^+$ and
$p\phi$ systems, we established upper limits on the cross section ratios
$\sigma(\Theta^+\bar{K}^0)~/~\sigma(\Lambda(1520)K^+)~<$~0.02 
and
$\sigma(\Theta^+\bar{K}^0)~/~\sigma(p\phi~)<$~0.15
at 90\% CL and a preliminary upper limit for the forward hemisphere cross 
section 
$\sigma(\Theta^+\bar{K}^0)~<$~30 nb/nucleon.
} 
\maketitle
\section{Introduction}
\label{intro}
Although the list of the experiments supporting the first  
observation~\cite{Nakano:2003qx} of the narrow $\Theta^+$ baryon 
with exotic quantum numbers is impressive~[2-13],  
its properties (spin, parity, total width and production cross sections) 
are not established and even its mere existence
is far from being proved.
The experimental concerns about the $\Theta^+$ baryon include the
statistical significance of the observed peaks, evident discrepancy
in its mass measured in $nK^+$ and $pK^0_S$ decay modes, the physical
meaning of 
cuts used to enrich and even to see the signal. Possible sources of the
false peaks were proposed for  $nK^+$ final 
state~\cite{Dzierba:2003cm,Rosner:2003ia} as well as for 
$pK^0_S$ one~\cite{Zavertyaev:2003wv}. The discussion of these questions
can be found in \cite{Zhao:2004rg,Klempt:2004yz}.

Recently negative results in a search for  $\Theta^+/\bar{\Theta}^-$ 
started to 
emerge~\cite{BES,HERA-B,PHENIX,ALEPH,DELPHI}, 
some of them still having the 
status of conference presentations. However, negative results
come either from the study of specific processes ($J/\Psi$ and $Z$ decays)
or from high energy inclusive nucleon/nuclei interactions, whereas
the positive ones come from $\gamma/K^+/\nu/e$ beams or relatively
low energy proton reactions. A possible explanation why the $\Theta^+$
has been seen in some experiments and not in the others was proposed 
in~\cite{Karliner:2004gr}.
In any case it is evident that more experiments are needed with different 
beams, targets, energies and higher statistics to reject or confirm the 
existence of the $\Theta^+$ baryon and establish its properties.

SPHINX experiment has a long history of searches for exotic baryons
and other exotic structures in various proton induced exclusive 
and semi-inclusive reactions.
As a result of the first stage of the experiment, we
published 
upper limits for the productions of heavy ($M>2.3$~GeV)
narrow states in the $p\phi$, $\Lambda(1520)K^+$, $pK^+K^-$ 
and $\Sigma(1385)^0K^+$ systems~\cite{Balats:zf} as well as in 
the $p \bar{p}$ and $p p \bar{p}$ systems~\cite{Vavilov:1993cv}.
Searches for the narrow $N_{\phi}(1960)$ baryon were also 
negative~\cite{Balats:xh}.
At the same time  we found an interesting structures in $\Sigma^0K^+$, 
$\Sigma^+\bar{K}^0$, $\Sigma(1385)^0K^+$ and $p\eta$ systems 
in the 1.7-2.1~GeV mass region. The origin of these structures is uncertain,
they need further study and confirmation in other experiments. 
The status of the study of some of these structures can be found 
in~\cite{Golovkin:1999ww,Vavilov:2000vw,Landsberg:1999gr,Antipov:eh}. 

The search for the $\Theta^+$ at SPHINX has its own history. After the
publication of the paper by 
Diakonov, Petrov and Polyakov~\cite{Diakonov:1997mm} in 1997, 
the planning have
started and hadron calorimeter has been installed in 1998 
with the idea to have additional capabilities to detect neutral hadrons.
However, we always had in mind that in a similar model 
Weigel~\cite{Weigel:1998vt} predicted much heavier (1580 vs 1530 MeV)
and much wider (100 vs $\lesssim$15 MeV) exotic baryon. The relatively
big width of the Weigel state could demand the background subtraction.
After a lengthy reconstruction of more than 600 millions events
recorded in 1998-1999, in 2001-2002 we made a first attempt to find
the $\Theta^+$ baryon using the final states without neutral hadrons.
We looked at the reaction\footnote
{Here and below we are using $N$ instead of C($N$) in the reaction
notation. We do not distinguish the processes on C nuclei and
quasifree nucleons in this work. However, they can be easily
separated using $P_T^2$ distribution.
}
\begin{equation}
\label{re-1}
p + N \to \Theta^+ \bar{K}^{*0} + N;\,\Theta^+\to pK_S^0,\,
\bar{K}^{*0}\to K^-\pi^+ 
\end{equation}
where $\Theta^+$ can be produced and compared it to the reaction
\begin{equation}
\label{re-2}
p + N \to pK_S^0 K^{*0} + N;\, K^{*0}\to K^+\pi^-
\end{equation}
which can be used to estimate the background.
Surprisingly enough, a narrow peak at $M=1548$~MeV was found in the
signal but not in the background reaction. However, it had a low 
significance (3-3.5 $\sigma$) and 
was found to be unstable against cuts.
Later on, with a better understanding of the neutron reconstruction,
we made a quick look at the same reaction with different final state
\begin{equation}
\label{re-3}
p + N \to \Theta^+ \bar{K}^{*0} + N;\, \Theta^+ \to nK^+,
\,\bar{K}^{*0} \to K^-\pi^+
\end{equation}
and did not find expected signal. Currently we consider an early
peak to be a normal statistical fluctuation. 
The reactions~(\ref{re-1})--(\ref{re-3})
are still under study and the results will be available in the near future.

In this work we present results of a search for $\Theta^+$ baryon
in more simple reaction
\begin{equation}
\label{re-4}
p + N  \to \Theta^+ \bar{K}^{0} + N
\end{equation}
The basic idea of our approach is to study simultaneously all
experimentally available final states of the $\Theta^+\bar{K}^0$ system,
thus eliminating the influence of possible reflections and inevitable
statistical fluctuations on the final judgment. 

\section{Experimental apparatus}
\label{sec-setup}
\begin{figure*}
\includegraphics[width=1.00\textwidth]{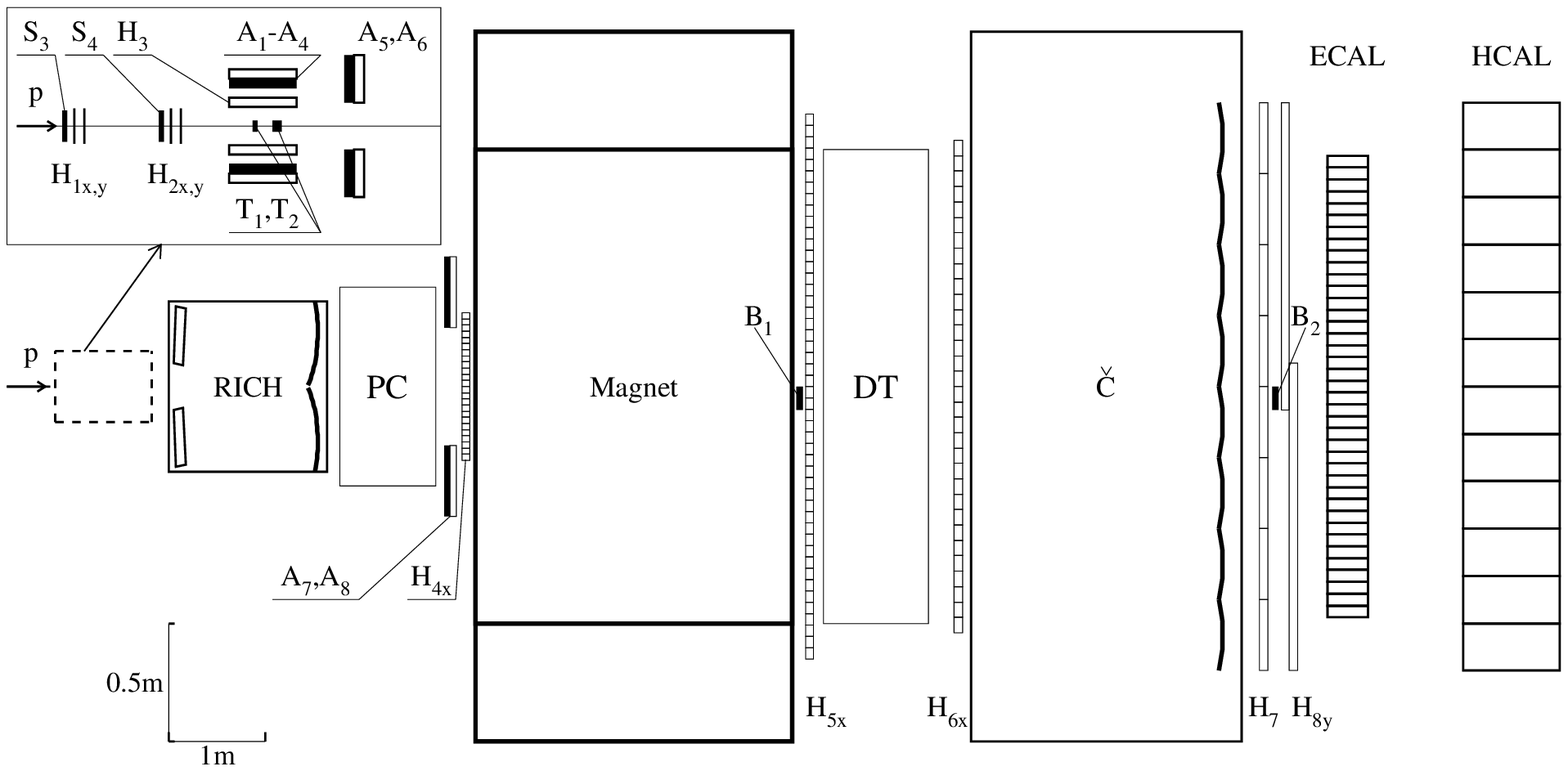}
\caption{%
Layout of the SPHINX spectrometer:
$\mathrm{S}_1$--$\mathrm{S}_4$~---
beam scintillator counters
(the very upstream counters $\mathrm{S}_1$ and $\mathrm{S}_2$ not shown in this
Fig.);
$\mathrm{T}_1$,~$\mathrm{T}_2$~---
copper and carbon targets;
$\mathrm{A}_1$--$\mathrm{A}_8$~---
lead-scintillator veto counters;
$\mathrm{B}_1$--$\mathrm{B}_2$~---
noninterected primary beam veto telescope;
$\mathrm{H}_{1X,Y}$, $\mathrm{H}_{2X,Y}$~--- beam hodoscopes;
$\mathrm{H}_3$~--- side hodoscope around the target;
$\mathrm{H}_{4X}$, $\mathrm{H}_{5X}$, $\mathrm{H}_{6X}$, $\mathrm{H}_{7}$, 
$\mathrm{H}_{8Y}$~--- trigger hodoscopes;
PC~--- proportional chambers;
DT~--- drift tubes;
RICH~--- velocity spectrometer with registration of the rings 
of Cherenkov radiation;
\v{C}~--- multichannel threshold Cherenkov counter;
ECAL~--- electromagnetic calorimeter;
HCAL~--- hadron calorimeter (for details see text).
}
\label{fig-sphinx21}
\end{figure*}

The SPHINX spectrometer was running in the proton beam of the IHEP accelerator 
with energy
$E_p=70$~GeV and intensity $I\simeq (2-4)\times 10^6$ p/spill in 1989--1999.
During that time several modifications of the detector were made.
The  data presented in this work was obtained with the last completely 
upgraded version of the
SPHINX spectrometer \cite{Antipov:eh}. The layout of this detector
is  presented in Fig.~\ref{fig-sphinx21}. 
The right-handed $X,Y,Z$ coordinate system of the setup had $Z$-axis in 
the direction of the proton beam, vertical $Y$-axis and horizontal 
$X$-axis. The origin of the coordinate system was in the center of the Magnet. 
The main elements of the detector are as follows:
\begin{itemize}
\item[1.] Detectors of the primary proton beam~--- scintillation counters 
$S_1 - S_4$ and scintillation hodoscopes $H_{1X,Y}$ and $H_{2X,Y}$.
\item[2.] The targets $T_1$ (Cu; 2.64~g/cm$^2$) and $T_2$ (C; 11.3~g/cm$^2$),
which were exposed simultaneously. 
The distance between the targets was 25~cm. 
The counter system around the target region included scintillation hodoscope 
$H_3$ (four counters per quadrant)
and veto counters~--- lead-scintillator sandwiches $A_1 - A_4$ 
around the targets and $A_5 - A_8$ in the forward direction. The holes 
in the   counters $A_5, A_6$ were  matched with the acceptance of 
spectrometer.
\item[3.] Wide-aperture magnetic spectrometer, based on the upgraded magnet 
SP-40 with uniform magnetic field in the volume 
of $100\times 70\times 150$~cm$^3$ and $p_T=0.588$~GeV/$c$, was equipped with 
proportional chambers PC, drift tubes DT and hodoscopes 
$H_{4X},H_{5X},H_{6X},H_{7},H_{8Y}$.
PC system consisted of five X and five Y planes with 2~mm pitch and 
sensitive region of $76.8\times 64.0$~cm$^2$.
DT system contained 18 planes of the thin-walled mylar tubes with the 
diameter of 6.25~cm. There were 32 tubes in each plane with the wires 
at $0^{\circ};\,\pm 7.5^{\circ}$ with respect to the vertical $Y$-axis. 
The sensitivity of the central tube of each plane in the beam region was 
artificially reduced. 
Electric field distribution in the drift tubes was described
in~\cite{Antipov:1994fn}.
The space resolution of the DT plane was $\simeq 300\,\mu$m on average. 
\item[4.] The system of Cherenkov counters for identification of secondary 
particles included RICH velocity spectrometer with photomatrix with 736 
small phototubes~\cite{Balats:zf, Kozhevnikov:rv}
and hodoscopical threshold Cherenkov counter~\v{C} with 32 optically 
independent channels. RICH detector was filled with 
SF$_6$ at a pressure slightly above atmospheric one. The threshold momenta
of $\pi/K/p$ in this detector were 3.5/12.4/23.6~GeV/$c$ correspondingly. 
\v{C} detector with an air at atmospheric 
pressure had momentum thresholds 6.0/21.3/40.1~GeV/$c$. 
Optical cells of \v{C} matched 
geometrically corresponding cells of  matrix hodoscope $H_7$ and 
this system can be used for charged particles identification at trigger 
level. 
\item[5.] Lead glass electromagnetic calorimeter ECAL was a 
matrix with $39\times 27$ cells of $5\times 5$~cm$^2$ each. This calorimeter 
was 
used previously in EHS experiment at CERN~\cite{Powell:1981it}. One  
counter was removed for the proton beam to pass through the ECAL.
\item[6.] Hadron calorimeter HCAL was a matrix of $12\times 8$ 
steel/scintillator $20\times 20$~cm$^2$ total absorption counters 
($5L_{\mathrm{abs}}$ thickness)~\cite{Antipov:1989vm}.
\end{itemize}
The beam was produced diffractively off the main beam of the U-70 accelerator
and had negligible momentum spread, small space dimensions 
($2{\times}4\,\mathrm{mm}^2$)
and small angular divergence. This allowed an effective use of the small 
$\mathrm{B}_1$--$\mathrm{B}_2$ counters as a beam killer telescope.
A special trigger logic scheme allowed the construction of up to eight
different kinds of triggers, using as a primitive elements  signals from
scintillation and veto counters, multiplicity of hits in the hodoscopes and  
threshold Cherenkov counter, and total energy sum in ECAL. 

The data acquisition system was based on MISS 
standard~\cite{Bityukov:1994ij} developed 
at IHEP and could record up to 4000 events per 10 second spill of 
the accelerator. 

The statistics used in this work was written during the last run of the 
SPHINX experiment in March-April of 1999. More than 600 million trigger
events were written, corresponding to $3{\cdot}10^{11}$ live protons on 
targets.
\section{Data analysis}
\label{sec-anal}
\subsection{General considerations}
\label{ssec-anal-gen}
The $\Theta^+\bar{K}^0$-system in the reaction~(\ref{re-4}) can have the final 
states presented in Table~\ref{tab-1}. 
Typical energy of $K_L^0$ for the processes under study is $\sim 15$~GeV. 
The decay length of $K_L^0$ was more than 450 metres on average, thus  it 
can be 
regarded as a stable
particle and reconstructed as a neutral 
cluster in ECAL or NCAL in the same way as a neutron. 
\begin{table*}
\begin{center}
\begin{tabular}{|cccc|c|c|c|}
\hline\noalign{\smallskip}
\multicolumn{4}{|c|}{Physical states and branching ratio}  & Experimental &      Total  branching     & Available \\ 
1 &    1/2        &   1/4         &   1/8        & final state &  ratio      & at SPHINX \\ 
\noalign{\smallskip}\hline\noalign{\smallskip}
 &               & $[nK^+]K_S^0$   &             & $nK^+\pi^+\pi^-$& (1/4)${*}$0.69 & Yes  \\
 & $\Theta^+K_S^0$ &               &             & &          &    \\
 &               &               & $[pK_S^0]K_S^0$ & $p\pi^+\pi^-\pi^+\pi^-$& (1/8)${*}$0.47 & Yes  \\
 &               & $[pK^0]K_S^0$   &             & &          &   \\
 &               &               & $[pK_L^0]K_S^0$ & $pK_L^0\pi^+\pi^-$& (1/8)${*}$0.69 & Yes  \\
$\Theta^+\bar{K^0}$ &               &               &             & &          &    \\
 &               &               & $[pK_S^0]K_L^0$ & $pK_L^0\pi^+\pi^-$& (1/8)${*}$0.69 & Yes  \\
 &               & $[pK^0]K_L^0$   &             & &          &    \\
 &               &               & $[pK_L^0]K_L^0$ & & 1/8      &  No  \\
 & $\Theta^+K_L^0$ &               &             & &          &    \\
 &               & $[nK^+]K_L^0$   &             & & 1/4      &  No   \\
\noalign{\smallskip}\hline
\end{tabular}
\end{center}
\caption{%
Possible final states of the $\Theta^+\bar{K}^0$ system in the 
reaction~(\ref{re-4}). It is assumed that the $\Theta^+$ baryon has only two 
modes 
of decay with equal probabilities. Only $\pi^+\pi^-$ decays of $K^0_S$ meson
are considered. 
}
\label{tab-1}
\end{table*}
The last two final states in Table~\ref{tab-1} have only one charged 
track and are unavailable at SPHINX due to trigger restrictions, while
the others were registered by three and five track triggers, which had
the following structure:
\begin{equation}
\label{trig-3}
T_{(3)} = T_0*H_3(0-1)*H_4(2-3)*H_6(\equiv 3)*H_7(1-3)
\end{equation}
\begin{equation}
\label{trig-5}
T_{(5)} = T_0*H_3(0-1)*H_4(4-5)*H_6(\equiv 5)*H_7(3-6)
\end{equation}
where pretrigger $T_0~=~S_1S_2S_3S_4*(\overline{B_1B_2})*\bar{A}_{5-8}$
and $H_i(m_1-m_2)$ means multiplicity requirement between $m_1$ and
$m_2$ for number of hits in  
hodoscope $H_i$. For technical reasons five track trigger was 
implemented not from the very beginning of the run, resulting in a lower 
luminosity for this kind of trigger.
In addition a simple beam trigger $T_{\mathrm{beam}}$, which was a greatly
prescaled ($\approx 16000$) fourfold coincidence $S_1S_2S_3S_4$, was written
throughout the run.

Four experimentally available final states of the $\Theta^+\bar{K^0}$ initial
state have different and to some extent complementary properties, which are
(together with the simulated characteristics of SPHINX detector) summarized
below:
\begin{itemize}
\item
$[nK^+]K_S^0$: 
Definite strangeness, large branching ratio, moderate effective
mass resolution, $\Lambda(1520)\to nK_S^0$ calibration decay in the same
final state, but: neutron should be detected, background from $\Lambda(1520)$. 
\item
$[pK_S^0]K_L^0$:
The best effective mass resolution, moderate branching ratio, 
$\phi\to K_S^0K_L^0$ 
calibration decay in the same final state, but: indefinite strangeness,
$K_L^0$ should be detected, background from $\phi$.
\item
$[pK_L^0]K_S^0$:
Moderate branching ratio, $\phi\to K_S^0K_L^0$ 
calibration decay in the same final state, but: indefinite strangeness,
worst effective mass resolution, $K_L^0$ should be detected, 
background from $\phi$.
\item
$[pK_S^0]K_S^0$:
No neutral particles in the final state (thus an additional constraint on 
the total energy in the event), moderate effective mass resolution, no
or small background from the known particles, but: indefinite strangeness,
small branching ratio and
small efficiency (five tracks, two weak decays), lower luminosity. 
\end{itemize}
For all but $[pK_S^0]K_S^0$ final state, a neutral hadron ($K_L$ or $n$) 
should be reconstructed. In the SPHINX detector approximately 65\% of neutrons 
and somewhat lower fraction of $K_L^0$ interact in ECAL, giving highly
fluctuating hadron showers with a typical visual energy release 
$\sim 20\%$ of their total energy. All the others 
are totally absorbed in NCAL except for a small fraction 
($\sim 5\%$), interacting in dead material in between. 
For neutral hadrons interacting in ECAL, only coordinates can be measured, 
the energy of hadron should be calculated using the exclusivity of the 
event. On the contrary, for neutral hadrons in NCAL both coordinates and 
energy can be measured in one detector, giving additional constraint on 
the exclusivity of the event. However, it was found by the Monte-Carlo
simulation and verified with the study of $\Lambda(1520)\to nK_S^0$ 
decays, that
the method of neutral hadrons reconstruction in ECAL 
gives better effective mass resolution for $nK^+$ and $pK_L^0$ systems
due to the excellent ability of the SPHINX setup to isolate exclusive 
processes 
and negligible momentum spread of diffractively produced proton beam.
In addition, the sample of events with ECAL neutral hadrons is
more than two times bigger.
Only the sample of events with $n(K_L^0)$ in interacting
ECAL is used in this work.
The sample of events with $n(K_L^0)$ in NCAL was our strategic reserve
and was planned to be used as a control one in the case of $\Theta^+$
observation. 

Being products of the decay of the same initial state, the 
experimental
final states~(Table~\ref{tab-1}) have different final particle sets 
and even different multiplicities. In order to have conclusive results,
a careful relative normalization and calibration is needed for
all reactions under study, including the reactions selected by
different triggers. 
It was provided by studying reactions 
\begin{equation}
\label{re-5}
p + N  \to \Lambda(1520)K^+ + N
\end{equation}
and
\begin{equation}
\label{re-6}
p + N  \to p\phi + N
\end{equation}
and then using different decay modes of $\phi$-meson and especially
those of $\Lambda(1520)$-hyperon, which has a lot of well 
measured decay modes~\cite{Hagiwara:fs} with quite different topologies
and multiplicities. With all this in mind, the 
following strategy was adopted:
\begin{itemize}
\item 
STEP 1:
Use beam trigger (minibias events) to investigate the ability of the setup
to isolate exclusive processes, the ability of the MC simulation to 
reproduce trigger conditions for main types of trigger. High 
intensity
reactions like $p+N \to p\pi^+\pi^- +N$ and 
$p+N\to p\pi^+\pi^-\pi^0 +N$ can be used for this study.
\item
STEP 2:
Develop Monte-Carlo generators for the simulation of exclusive production
of $\Lambda(1520)K^+$ and $p\phi$ systems 
(reactions~(\ref{re-5})--(\ref{re-6}))
and adjust them to
reproduce the experimentally measured kinematics for decay modes
with large statistics, $\Lambda(1520)\to pK^-$ (21k events) and
$\phi\to K^+K^-$ (10k events).
\item
STEP 3:
Understand the ability of the setup to isolate and reconstruct  
exclusive reactions with neutral hadrons ($n$, $K_L^0$) in the final state.
It can be done comparing the generated and reconstructed reactions
with $\Lambda(1520)\to nK_S^0$ and $\phi\to K_S^0K_L^0$ decays.
\item
STEP 4: 
Understand the ability of the setup to isolate and reconstruct the 
exclusive reactions with five tracks in the final state using the
decay $\Lambda(1520) \to \Lambda\pi^+\pi^-$.
\item
STEP 5: In addition, verify the ability of the setup to reconstruct 
as wide set of the
different topologies of the final state as possible by using generated and
reconstructed decays of $\Lambda(1520)$ to $ \Sigma^+\pi^-,\Sigma^0\pi^0$ and
 $\Sigma^-\pi^+$.
\end{itemize}
In other words, the plan was to study the whole set of calibration reactions
\begin{equation}
\label{re-calib}
\begin{array}{rcll}
p + N & \to & p\pi^+\pi^-      &  +\, N \\
p + N       & \to & p\pi^+\pi^-\pi^0 & +\, N  \\
p + N       & \to & pK^+K^-          & +\, N  \\
p + N       & \to & [nK_S^0]K^+        & +\, N   \\
p + N       & \to & p[K_S^0K_L^0]      & +\, N  \\
p + N       & \to & [\Lambda\pi^+\pi^-]K^+          & +\, N  \\
p + N       & \to & [\Sigma^+\pi^-]K^+          & +\, N  \\
p + N       & \to & [\Sigma^-\pi^+]K^+          & +\, N  \\
p + N       & \to & [\Sigma^0\pi^0]K^+          & +\, N  
\end{array}
\end{equation}
for different purposes and in a certain order.
Only after Steps 1--5 are successfully done, we should start to look 
for $\Theta^+$ in the set of signal reactions
\begin{equation}
\label{re-signal}
\begin{array}{rcll}
p + N & \to & [nK^+]K_S^0        &  +\, N \\
p + N       & \to & [pK_S^0]K_L^0      & +\, N   \\
p + N       & \to & [pK_L^0]K_S^0      & +\, N   \\
p + N       & \to & [pK_S^0]K_S^0      & +\, N   
\end{array}
\end{equation}
some of them being the part of the calibration set. 
If existent and produced in the process under study,
$\Theta^+$ should emerge simultaneously in all four final states 
in accordance with their relative probabilities and efficiencies.

\subsection{Study of calibration reactions}
\label{ssec-anal-calib}
\begin{figure}
\includegraphics[width=\hsize]{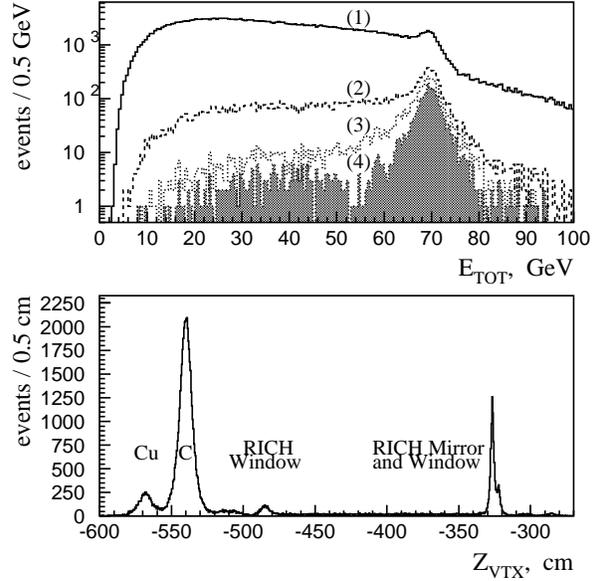}
\caption{%
Events from the beam trigger. Upper panel: Total energy of secondary 
tracks for (1)-all two positive/one negative track events,
(2)-with trigger ($T_{(3)}$) simulation cuts, (3)-no neutral clusters in ECAL,
(4)-good primary vertex in targets. Lower panel: $z$-coordinate of
good primary vertex for case (3) of the upper panel.
}
\label{fig-beam}
\end{figure}

The first step in the analysis is illustrated by Fig~\ref{fig-beam}.
All $T_{\mathrm{beam}}$ events were reconstructed with a standard tracking 
program
requiring at least one track after the magnet. The events
with exactly 
two positive and one negative track
after the magnet were then selected, and
this sample of unbiased events was used to study the efficiency of 
trigger elements and trigger itself ($T_{(3)}$ in this case).
The peak in total energy at 70~GeV corresponds
to exclusive events mainly of the $p\pi^+\pi^-$ type and inelastic background
after imposing trigger simulation cuts and simple additional ones is
quite small. This figure also demonstrates that interactions in targets
can be easily isolated with a small background. Only events
with a primary vertex in Carbon target were used in further study. The sample of 
events from Copper target was again left as a control one.

\begin{figure}
\includegraphics[width=\hsize]{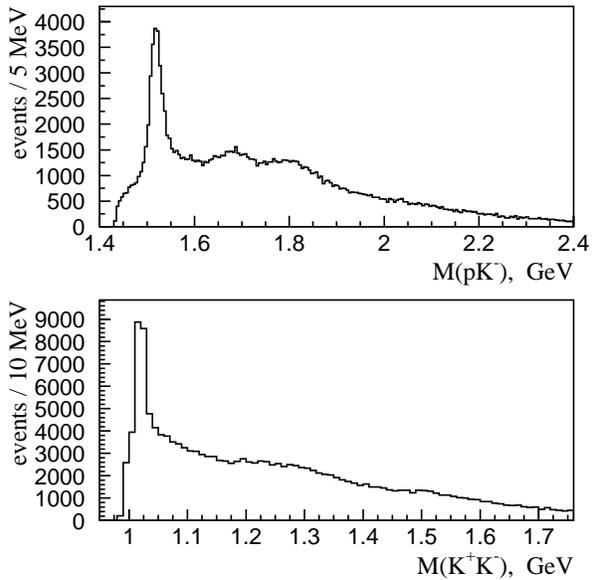}
\caption{%
Effective mass distributions $M(pK^-)$ and $M(K^+K^-)$ for the reaction
$p+N \to pK^+K^- +N$ after applying selection cuts described in the text. 
}
\label{fig-l1520-phi}
\end{figure}

To study the production characteristics of  
reactions~(\ref{re-5}) and (\ref{re-6}), which was the subject of the second 
step, the events of the reaction $p+N \to [pK^+K^-] +N$ were selected
from $T_{(3)}$ trigger sample using the following criteria:
\begin{itemize}
\item Two positive and one negative track after the magnet;
\item Good primary vertex in the carbon target;
\item No neutral clusters with $E>1$~GeV in ECAL;
\item Energy balance, $65 < E_{\mathrm{tot}} = E_1+E_2+E_3 < 75$~GeV;
\item The momentum of any secondary particle $>5$~GeV/$c$;
\item Identification in RICH as a $pK^+K^-$ system.
\end{itemize}
The procedure of the identification of the final state by RICH detector
was described in our previous publications (see~\cite{Kozhevnikov:rv} 
for details)
and will be not discussed here. More than 160000 events passed the selections
cuts and results are presented in Fig~\ref{fig-l1520-phi}. The peaks
from $\Lambda(1520)\to pK^-$ and $\phi\to K^+K^-$ decays can be clearly
seen together with other well known structures.


The events from $\Lambda(1520)$ and $\phi$ peaks after background subtraction
were used to study the production characteristics of the $\Lambda(1520)K^+$
and $p\phi$ systems and to develop Monte-Carlo generators for the
simulation of the reactions~(\ref{re-5}) and (\ref{re-6}).

The simulation of signal and normalization processes in
the SPHINX setup was done in the framework of GEANT 3.21 package.
This included in particular:
\begin{itemize}
\item Detailed description of geometry and material of the setup including,
for example, individual wires in drift tubes;
\item Realistic simulation of the efficiencies of trigger elements; 
\item Simulation of experimentally measured inefficiency of tracking devices 
      (PC and DT) including inefficiency in the beam region;
\item Propagation of all secondary particles generated by GEANT including
electromagnetic and hadron showers in the $\gamma$-spectrometer;
\item Simulation of the multibeam events with more than one beam particle
within $\pm 600$~ns time window of the main process and then realistic
simulation of signal development in various detectors.
\end{itemize}
The simulation of the events was done up to the level of digitized 
detector responses so that they were processed in exactly the same way as real 
events.

\begin{figure}
\includegraphics[width=\hsize]{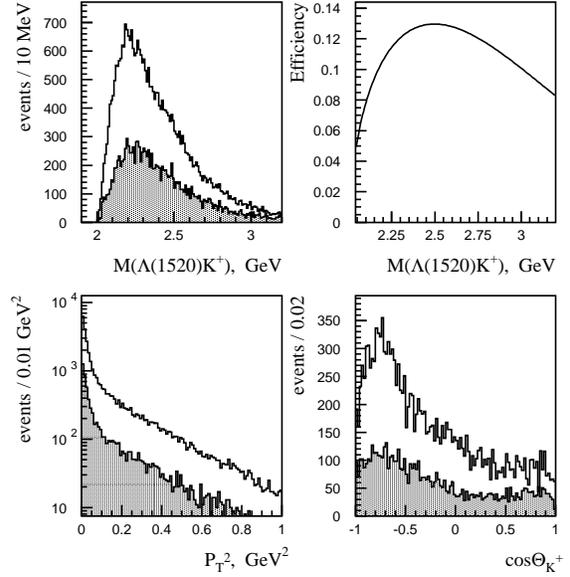}
\caption{%
A comparison of reconstructed and simulated distributions
for the reaction $p + N \to \Lambda(1520)K^+ +N$ with 
$\Lambda(1520)\to pK^-$ decay. Open histogram represents data,
hashed - MC. Only a small fraction of MC events is used in this
picture. Also shown is the efficiency.
}
\label{fig-l1520-prod}
\end{figure}

After a few iterations it was possible to reproduce the experimentally 
observed characteristics of $\Lambda(1520)K^+$ and $p\phi$ production
with a reasonable accuracy. 
The variables studied were: the mass $M$ of the system 
$\Lambda(1520)K^+\,(p\phi)$, the transverse momentum squared $P_T^2$ and
the angles $(\theta^*,\phi^*)$ of the $\Lambda(1520)\,(\phi)$ in the
$\Lambda(1520)K^+\,(p\phi)$ rest frame (Gottfried-Jackson reference frame).
The comparison of experimental and Monte-Carlo
distributions (after passing the analysis chain) is presented
in Fig.~\ref{fig-l1520-prod} for $\Lambda(1520)K^+$ production. The
distributions for $p\phi$ system are quite similar, as it was 
observed already in our earlier work~\cite{Balats:zf}. Both systems exhibit
low mass enhancement and distinct evidence for the coherent production
on the Carbon nuclei (see $P_T^2$ distribution). The most significant
difference between two systems is a $\cos\theta^*$ distribution, 
which for $p\phi$ is more flat. The detailed study showed, however,
that the efficiency for the processes under study is quite insensitive
to the $P_T^2$ (up to $\approx2~\mathrm{GeV}^2/c^2$) and production angles 
of the 
resulting $NK\bar{K}$ system. For example, for flat $\cos\theta^*$ distribution
for the production of $\Lambda(1520)K^+$ system, the overall efficiency 
only changes from 10.5\% to 11\%. 

With generators at hand, it was then possible to simulate 
other decay channels of $\Lambda(1520)$ and $\phi$ and to compare
the results of the simulation with experimental distributions.

\begin{figure}
\includegraphics[width=\hsize]{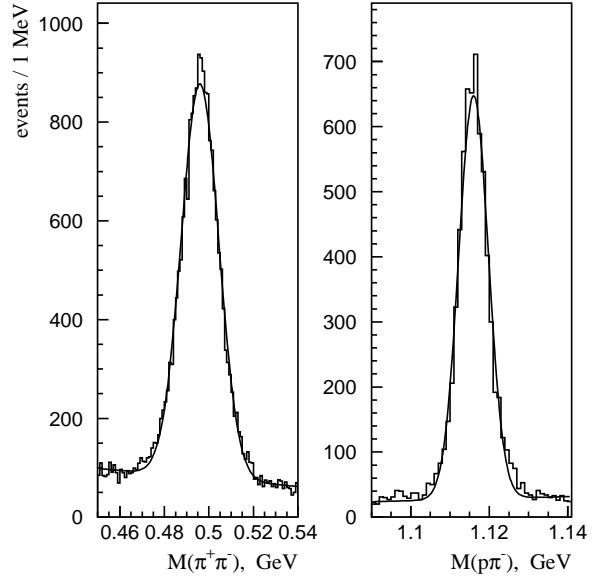}
\caption{%
Signals from decays of $K_S^0\to\pi^+\pi^-$ and
$\Lambda\to p\pi^-$ in the reactions $p + N \to nK_S^0K^+ + N$
and $p + N \to \Lambda\pi^+\pi^-K^+ + N$. The fit
gives: $M(K_S^0) = 496$~MeV, $\sigma(K_S^0) = 8.4$~MeV, 
$M(\Lambda) = 1116$~MeV, $\sigma(\Lambda) = 3.8$~MeV. 
}
\label{fig-ks-lam}
\end{figure}

The events corresponding to the reactions
\begin{equation}
\label{re-nkskp}
p + N  \to nK_S^0K^+   + N
\end{equation}
were selected from three track sample as follows:
\begin{itemize}
\item Good secondary decay vertex lies in the allowed region: 
$-520\,< z_{\mathrm{sec}} < -270\,\mathrm{cm}$.
\item Good primary vertex, composed of the vee vector, unpaired secondary
track and beam track
 is in the Carbon target: $-555\, < z_{\mathrm{prim}} < -525$~cm.
\item $l/\sigma_z>3$, where $l=z_{\mathrm{sec}}-z_{\mathrm{prim}}$ and 
$\sigma_z$
is the calculated accuracy for this quantity.
\item Missing energy 
$E_{\mathrm{miss}}=E_{\mathrm{beam}}-E_{\pi^+}-E_{\pi^-}-E_{K^+}>5$~GeV. 
\item There is only one neutral cluster in ECAL with $E>1$~GeV.
\item The momentum of the unpaired positive track $>5$~GeV/$c$. 
\item RICH readings are consistent with the hypothesis that the negative
and the positive tracks forming the secondary vertex are pions and 
the remaining track is kaon.
\item The effective mass $M(\pi^+ \pi^-)$ is within $\pm 2.5\sigma$ of the
$K_S^0$ peak value.
\end{itemize}
The events corresponding to the reactions
\begin{equation}
\label{re-pkskl}
p + N  \to pK_S^0K_L^0   + N
\end{equation}
were selected in exactly the same way with the evident exchange 
kaon $\leftrightarrow$ proton in the identification requirement.
Approximately 20000 (30000) events of $nK_S^0K^+$ $(pK_S^0K_L^0)$ type were 
thus 
selected. Mass spectrum $\pi^+\pi^-$ for the $nK_S^0K^+$ system
before the final $K_S^0$ selection cut 
is shown in Fig.~\ref{fig-ks-lam}.

The energy of $n$ (or $K_L^0$, with the evident changes in formulas) 
is calculated as 
$E_{n}=E_{\mathrm{beam}}-E_{K_S^0}-E_{K^+}$ and direction of flight of 
neutron is calculated using coordinates of the ECAL cluster and primary 
vertex. The effective mass of  $nK_S^0$ system is then calculated using
the tabular value of $K_S^0$ mass as
\begin{equation}
\label{eq-1}
M(nK_S^0) = M(n\pi^+\pi^-) - M(\pi^+\pi^-) + M(K_S^0).
\end{equation}

\begin{figure}
\includegraphics[width=\hsize]{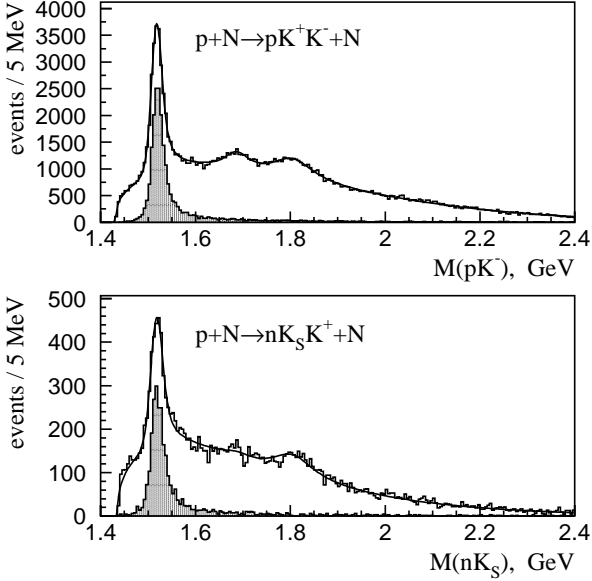}
\caption{%
Effective mass distributions $M(pK^-)$ and $M(nK_S^0)$ for the reaction
$p+N \to [N\bar{K}]K^+ +N$. MC-simulated signals for $\Lambda(1520)$ are shown 
hatched.  
}
\label{fig-l1520}
\end{figure}

\begin{figure}
\includegraphics[width=\hsize]{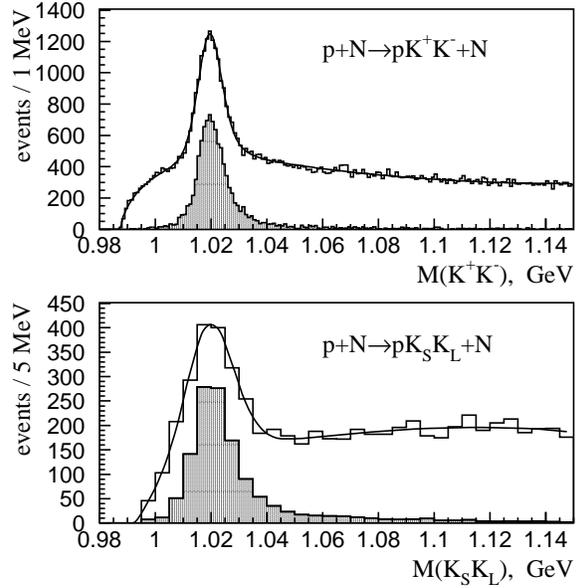}
\caption{%
Effective mass distributions $M(K^+K^-)$ and $M(K_S^0K_L^0)$ for the reaction
$p+N \to p[K\bar{K}] +N$. MC-simulated signals for $\phi$ are shown hatched. 
}
\label{fig-phi}
\end{figure}

In a threshold region this gives an improvement in the resolution by
20-30\%, in agreement with the MC simulation. 
The resulting mass spectra $M(nK_S^0)$ and $M(K_S^0K_L^0)$ are presented
in Fig.~\ref{fig-l1520} and~\ref{fig-phi}. We observe a good qualitative
agreement in the form of spectra for different final states
and also between resonance signals in the 
data and Monte-Carlo. 
The agreement is also good numerically, as will be shown later.


The reaction
\begin{equation}
p+N\to\Lambda(1520)K^++N;\, \Lambda(1520)\to\Lambda\pi^+\pi^- 
\label{re-lampipik}
\end{equation}
is a main calibration process for five track trigger. It allows to cross check
the simulation for three and five tracks events and connect $pK_S^0K_S^0$ 
decay mode of
the $\Theta^+\bar{K}^0$ system to all the others. The events were selected 
from five
track trigger sample as follows:
\begin{itemize}

\item Five secondary tracks with $\sum Q_i=1$.

\item No neutral clusters with $E>1$~GeV in ECAL.

\item The total energy of all five particles corresponds to the energy 
      of the incident proton: $65 < E_{\mathrm{tot}} < 75$~GeV.

\item Only one good secondary vertex in the allowed region. 

\item Good primary vertex composed of vee tracks and three others tracks
is in the Carbon target.
\item $l/\sigma_z>3$. 
\item RICH readings are consistent with the hypothesis that one of the 
unpaired positive tracks with momentum $>5$~GeV is a kaon. 
\item The effective mass of $p \pi^-$ combination is within $\pm 2.5\sigma$ of 
the $\Lambda$ peak value.
\end{itemize}
Events of $pK_S^0K_S^0$ final state were selected in a similar way, demanding 
two (instead of one) different secondary vertices and the lone
unpaired positive track identified as being consistent with proton
hypothesis. Both $\pi^+ \pi^-$ combinations were then required to be within 
$\pm 2.5\sigma$ of the $K_S^0$ peak value.

\begin{figure}
\includegraphics[width=\hsize]{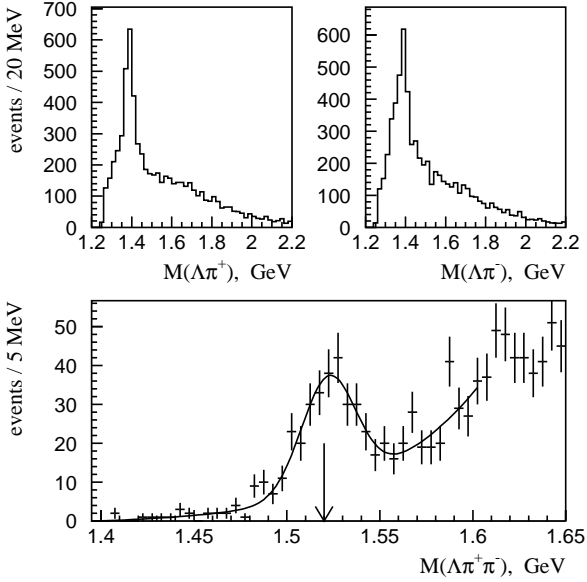}
\caption{%
Effective mass distributions $M(\Lambda\pi^+)$, $M(\Lambda\pi^-)$ and 
$M(\Lambda\pi^+\pi^-)$ for the reaction
$p+N \to \Lambda\pi^+\pi^-K^+ +N$. The peaks of $\Sigma(1385)^+$ and
$\Sigma(1385)^-$ are clearly seen (upper pictures). The arrow shows 
the nominal mass of $\Lambda(1520)$. 
}
\label{fig-lampipi}
\end{figure}

\begin{figure}
\includegraphics[width=\hsize]{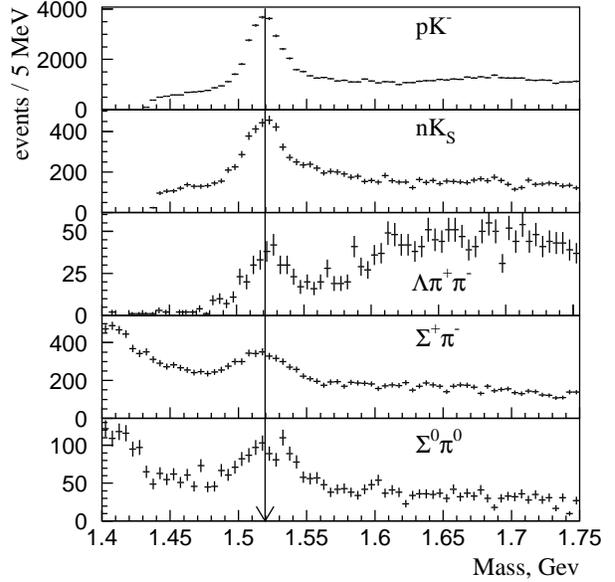}
\caption{%
Mass spectra for different decay modes of $\Lambda(1520)$ in
the reaction $p+N \to \Lambda(1520)K^+ +N$. The arrow shows the nominal
mass of $\Lambda(1520)$. 
}
\label{fig-l1520all}
\end{figure}

The quality of $\Lambda$ signal for $\Lambda\pi^+\pi^-K^+$ final state
can be seen in Fig.~\ref{fig-ks-lam} and mass spectra for some subsystems 
are shown in Fig~\ref{fig-lampipi}.
The reaction is dominated by the production of the 
$\Sigma(1385)^{\pm}\pi^{\mp}K^+$ system. The peak from 
$\Lambda(1520)$
in the $\Lambda\pi^+\pi^-$ effective mass spectrum is clearly seen and
the number of events is sufficient to make quantitative conclusions. 
Fig.~\ref{fig-l1520all} represents the summary of the results in the study of
$\Lambda(1520)K^+$ production in different decay modes. Two more decay modes 
are included, $\Lambda(1520)\to\Sigma^+\pi^-$ and 
$\Lambda(1520)\to\Sigma^0\pi^0$, with $\Sigma^+\to p\pi^0$ and 
$\Sigma^0\to \Lambda\gamma, \Lambda\to p\pi^-$ decays, corresponding to the
topologies ``kink+$2\gamma$'' and ``vee+$3\gamma$''. Details of the data 
processing for these modes will not be discussed here. This figure can be used
to estimate the systematics of the mass scale for quite different topologies
and final particle sets.

Up to now the results were presented at a qualitative level. Now we turn
to the discussion of the numerical results for calibration processes.
To estimate the number of events, the mass spectra in
Figs~\ref{fig-l1520} and \ref{fig-phi} were fitted by a sum of resonance and 
smooth background.
The peaks of $\Lambda(1520)$ and $\phi$ were described by relativistic
Breit-Wigner function with orbital momenta $L=2$ and $L=1$ 
smeared by gaussian resolution.
Widths of the resonances were fixed by world average values.
Two other peaks in the $pK^-$ mass spectrum are known to contain a
lot of $\Lambda^*/\Sigma^*$ states and were described by two simple
Breit-Wigner functions with free parameters for mass and width.
The background was chosen to have a form of
$P_1(\Delta M)^{P_2}\cdot\exp(-P_3\Delta M-P_4\Delta M^2)$ with free 
parameters $P_i$, $\Delta M = M - M_{\mathrm{thr}}$ and threshold mass 
$M_{\mathrm{thr}} = M_N + M_K$.
The resolution parameter of gaussian function was either free or fixed 
by MC value. The results of fits are presented in Table~\ref{tab-2}.
\begin{table*}[!t]
\begin{center}
\begin{tabular}{|cccccl|}
\hline\noalign{\smallskip}
 Particle       & Final state    & Events & $\varepsilon$, & Resolution &  $\sigma_{\mathrm{meas}}(\sigma_{\mathrm{corr}}$), \\
                &                &        & \%             & $\sigma$, MeV   & nb/nucleon  \\  
\noalign{\smallskip}\hline\noalign{\smallskip}
$\Lambda(1520)$ & $[pK^-]K^+$    & 21200$\pm$300           & 10.5     & 8.1(7.1)        &  1015(1400)        \\
                & $[nK_S^0]K^+$  & 2490$\pm$90             & 3.8      & 9.8(8.8)        &  ~965        \\
$\phi(1020)$    & $[K^+K^-]p$    & 10660$\pm$190           & 12.1     & 3.4(3.5)        &  ~202($279^*$)        \\
                & $[K_S^0K_L^0]p$& 1440$\pm$80             & 3.7      & 7.8(6.6)         &  ~188        \\ 
\noalign{\smallskip}\hline
\end{tabular}
\end{center}
\caption{%
Results of the fits of mass spectra in Figs.~\ref{fig-l1520} and \ref{fig-phi}.
Also the efficiencies (assuming $\mathrm{BR}(K_S^0\to\pi^+\pi^-) = 100\%$ for
the decay modes with $K_S^0$) and experimental(Monte-Carlo) 
resolution are shown. The number of events corresponds to fits with
experimental resolution.
For the meaning of cross section values see text below.  
}
\label{tab-2}
\end{table*}

The errors in the number of events are statistical.  
The systematical uncertainties include an uncertainty 
associated with the background description (polinomial function instead of
exponential with a smaller fitting range in mass), gaussian resolution
(experimental or MC)
and uncertainties in the PDG parameters of the $\Lambda(1520)$ 
and $\phi$ resonant width.
Added in quadrature, these do not exceed 4~(8)\% for the modes with 
high(low) statistics.

As a first check of the consistency of the results we can calculate the
relative branching for different decay modes. For $\Lambda(1520)$ we have
\begin{eqnarray}
\label{eq-lamratio}
\frac{\mathrm{BR}[\Lambda(1520)\to n\bar{K}^0]}
     {\mathrm{BR}[\Lambda(1520)\to pK^-]} = & & \nonumber \\  
 \hspace*{-6mm}=\frac{2\cdot N[\Lambda(1520)\to nK_S^0]/\mathrm{BR}[K_S^0\to \pi^+\pi^-]}
     {N[\Lambda(1520)\to pK^-]}\cdot 
\frac{\varepsilon_{pK^-}}{\varepsilon_{nK_S^0}} = & &\nonumber \\  
=\frac{2\cdot 2490/0.686}{21200}\cdot \frac{10.5}{3.8} = 0.94\pm0.08, & &
\end{eqnarray}
which is very close to the unity, as it should be. For $\phi$-meson decays
we have
\begin{eqnarray}
\label{eq-phiratio}
\frac{\mathrm{BR}[\phi\to K_S^0K_L^0]}
     {\mathrm{BR}[\phi\to K^+K^-]}  = & & \nonumber \\
 \hspace*{-1mm}=\frac{N[\phi\to K_S^0K_L^0]/\mathrm{BR}[K_S^0\to \pi^+\pi^-]}
     {N[\phi\to K^+K^-]}\cdot 
\frac{\varepsilon_{K^+K^-}}{\varepsilon_{K_S^0K_L^0}} = & & \nonumber \\  
 \hspace*{-30mm}=\frac{1440/0.686}{10660}\cdot \frac{12.1}{3.7} = 0.64\pm0.06, & &
\end{eqnarray}
which again is close to the tabular value $0.693\pm0.018$. 
We can conclude, therefore, that the processes with
neutron and $K_L^0$ in the final state can be reproduced at SPHINX
with the relative accuracy better than 10\%. This is also true for 
other decay modes of $\Lambda(1520)$ not shown in the table, including
the ``five track'' calibration decay $\Lambda(1520)\to\Lambda\pi^+\pi^-$,
however with a little bit worse accuracy. It should be noted, in addition,
that Monte-Carlo and experimental effective mass resolutions are close
to each other, the MC resolution being typically a little bit better.
This is important in a search for narrow states, where the estimations
rely heavily on the resolution. 

Another question of  major importance is the absolute calibration of
cross sections. In most of the models the cross section for the
production of the $\Theta^+$ baryon in a wide class of exclusive processes
is directly proportional to
it's width, being governed by the same constant $g^2_{\Theta NK}$.
Cross section for $\Lambda(1520)K^+$ and $p\phi$ production was calculated
as $\sigma = (1/L)\cdot N/(\mathrm{BR}\cdot\varepsilon)$, where 
luminosity $L$ per nucleon  was
estimated assuming $A^{2/3}$ dependence of the cross section on the 
mass number. The luminosity for Carbon target was found 
to be 884~events/nb for $T_{(3)}$ and 445~events/nb for $T_{(5)}$.   
The measured cross sections $\sigma_{\mathrm{meas}}$ 
can be found in Table~\ref{tab-2}. The data is preliminary, as not all
corrections, common to all processes and not influencing the relative
quantities like~(\ref{eq-lamratio}) and~(\ref{eq-phiratio}), were included 
into efficiency calculations. 
These corrections include, for example, rate dependent accidentals in veto
counters. In general they are small and are under study now.
Can we check the cross section values from 
independent measurements? As far as we know, there are no data on 
exclusive production of $\Lambda(1520)K^+$ system 
in nucleon-nucleon(nucleus) 
interactions and our result seems to be the first one of this kind. The
existing data for $\phi$-meson production in the reaction $pp\to \phi pp$
is scarce and do not allow the extrapolation to our energy. 
However, we can use 
the data for reaction $pp\to \omega pp$, which exist at higher as well as
lower energies. The analysis, done in~\cite{Arenton:ts}, allows to 
extrapolate to our energy, giving the value 
$(36\pm 4)/2 = 18\pm 2~\mu$b/nucleon for $\omega p$ forward
hemisphere production cross section. Using then  
cross section ratio 
$\sigma_{\phi}/\sigma_{\omega} = (1.55\pm 0.31)\times 10^{-2}$, measured in 
our study of OZI-rule~\cite{Golovkin:1997uu}, we arrived at the prediction for 
$\phi p$ production cross section $\sigma_{\phi p} = 279\pm 64$~nb/nucleon.
This value is close, but somewhat higher than $\sigma_{meas} = 202~$nb/nucleon 
(Table~\ref{tab-2}).
Being conservative, we used this (higher) cross section as input. It
is denoted by $(^*)$ sign in Table~\ref{tab-2}.
The correction factor $k_{\mathrm{corr}} = 279/202 = 1.38$ 
was used then to calculate  the corrected cross
section for $\Lambda(1520)K^+$ production.
in Table~\ref{tab-2}.
The same conservative correction factor will be used later in the estimations 
of absolute cross section for $\Theta^+$ production. 

\subsection{Study of signal reactions}
\label{ssec-anal-signal}
\begin{figure}
\includegraphics[width=\hsize]{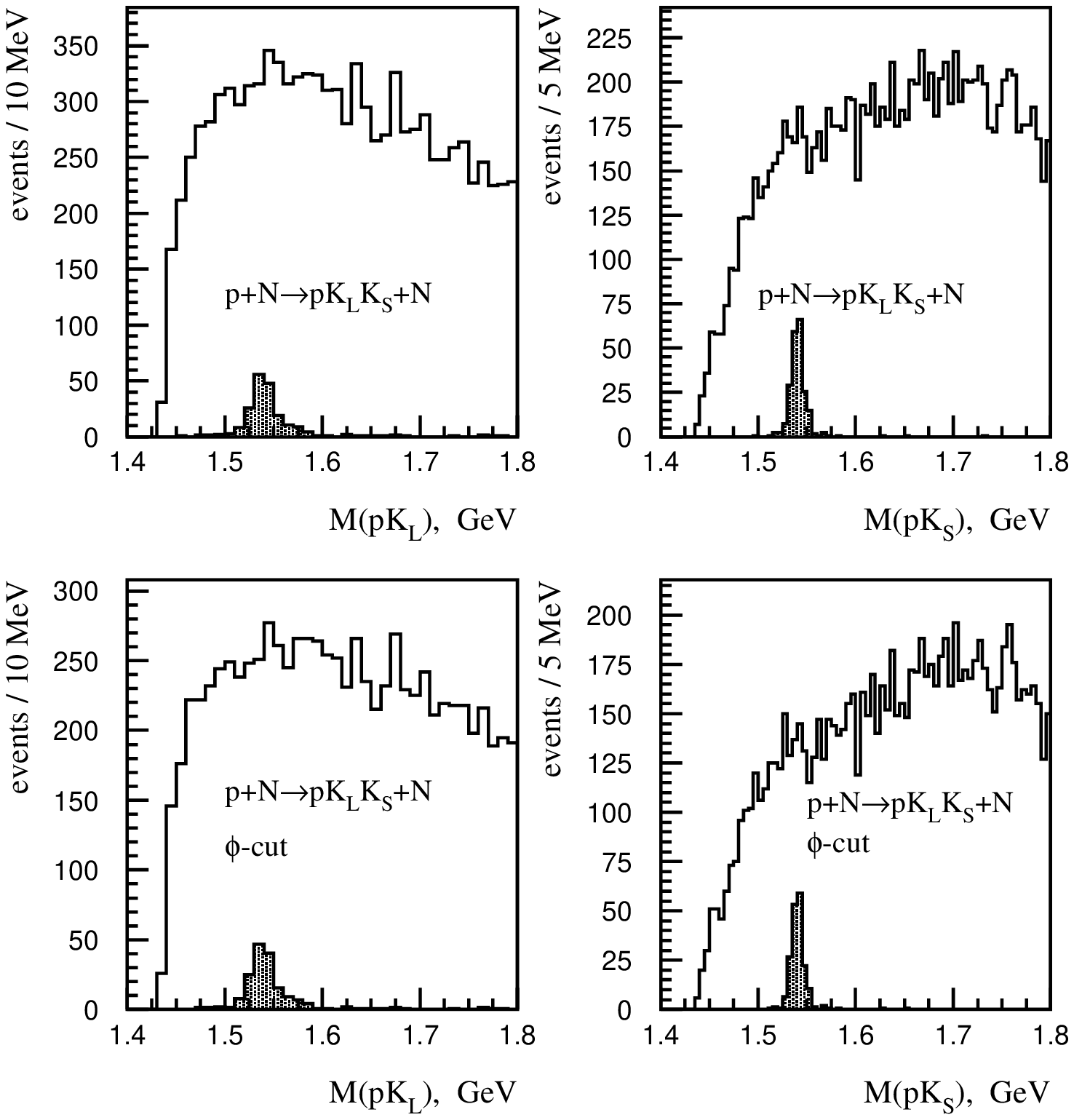}
\caption{%
Effective mass spectra of $pK_L^0$ and $pK_S^0$ systems in the reaction
$p+N \to pK_S^0K_L^0 +N$ with ($M(K_S^0K_L^0)>1.04$~GeV, lower row) and 
without ``$\phi-$cut'' (upper row).
MC-simulated signals (hatched) correspond to 
$\sigma_{\Theta^+ \bar{K^0}}/\sigma_{\Lambda(1520)K^+} = 0.1$.
}
\label{fig-pkskl}
\end{figure}

\begin{figure}
\includegraphics[width=\hsize]{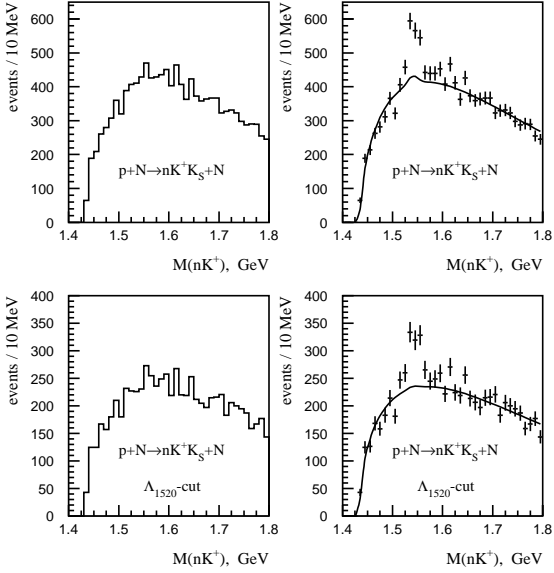}
\caption{%
Effective mass spectrum of $nK^+$ in the reaction
$p+N \to nK^+K_S^0 +N$ with ($M(nK_S^0)>1.55$~MeV, lower row) and 
without ``$\Lambda(1520)-$cut'' (upper row).
Pictures on the right panel are the sums
of real data and MC-simulated signals for 
$\sigma_{\Theta^+ \bar{K^0}}/\sigma_{\Lambda(1520)K^+} = 0.1$.
Curves in the right column are the results of the fit of left 
column distributions.
}
\label{fig-nkplus}
\end{figure}

The calibration reactions~(\ref{re-pkskl}) and (\ref{re-nkskp}) are at the same
time the reactions, where the $\Theta^+$ baryon can be searched for. 
Effective mass spectra $M(pK_L^0)$ and $M(pK_S^0)$ for the 
reaction~(\ref{re-pkskl})
are shown in Fig.~\ref{fig-pkskl}
and mass spectrum  $M(nK^+)$  
for the reaction~(\ref{re-nkskp}) in Fig.~\ref{fig-nkplus}. 
No evident structures can be seen in these distributions, except for a
hint on a shoulder in the $pK_S^0$ mass spectrum at a mass of 
${\approx}1510$~MeV.
However, this structure is completely absent in $M(pK_L^0)$ mass spectrum
as well as in $M(nK^+)$ mass spectrum.
Thus we don't see the signals for the $\Theta^+$ baryon and only upper
limits can be produced from these distributions.

We are searching for narrow signals and will assume that effective mass
distributions are completely dominated by experimental resolutions for each 
of the systems under study. This assumption is valid for
the width (FWHM) of the $\Theta^+$ baryon $<6-10$~MeV. The width like that 
and lower is indicated in most precise ``positive evidence'' experiments
and the assumption doesn't seem to be limiting. To achieve the limits on the 
number of events and then on production cross sections, the resolution
and efficiency for each of the final state is needed. These can be estimated
only by Monte-Carlo simulation. However, 
the production mechanism for $\Theta^+$ baryon is unknown, so some model 
is needed to simulate its production. 
We simulate the production of $\Theta^+K^0$ system as being similar to
that of $\Lambda(1520)K^+$ and $p\phi$ systems ($P_T^2$ and 
$M(\Theta^+\bar{K}^0$)) and 
the distributions of the decay angles of the $\Theta^+K^0$ system 
in Gottfried-Jackson frame were assumed to be isotropic. The assumptions
about production characteristics are not as limiting as they seem. 
The  efficiencies demonstrate very weak dependence on $P_T^2$ and
decay angles (see the discussion about the simulation of the 
$\Lambda(1520)K^+$ production above) and only if $\Theta^+$ prefers to
originate from the $\Theta^+K^0$ system with very big mass ($\geq$~3.5~GeV)
and very high $P_T$, the efficiency for the $\Theta^+K^0$ system detection
would be significantly overestimated. 

\begin{figure}
\includegraphics[width=\hsize]{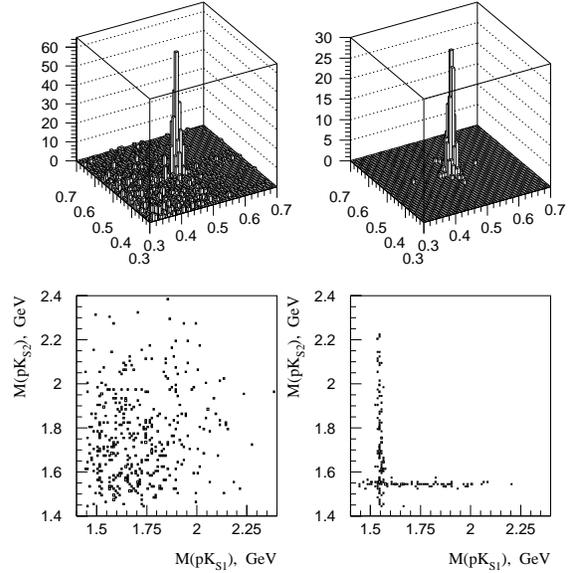}
\caption{%
Distributions for the reaction $p+N \to pK_S^0K_S^0 + N$.
Upper row --- $M(\pi_1^+\pi_2^-$) vs $M(\pi_3^+\pi_4^-)$ with all cuts 
but $K_S^0$ selection cut. 
Lower row --- $M(pK_{S2}^0$) vs $M(pK_{S1}^0)$ after all cuts.
Left column --- data, right column --- Monte-Carlo
simulation of reaction $p+N \to \Theta^+\bar{K}^0 + N,
~\Theta^+\bar{K}^0 \to [pK_S^0]K_S^0$. 
}
\label{fig-pksks-1}
\end{figure}

\begin{figure}
\includegraphics[width=\hsize]{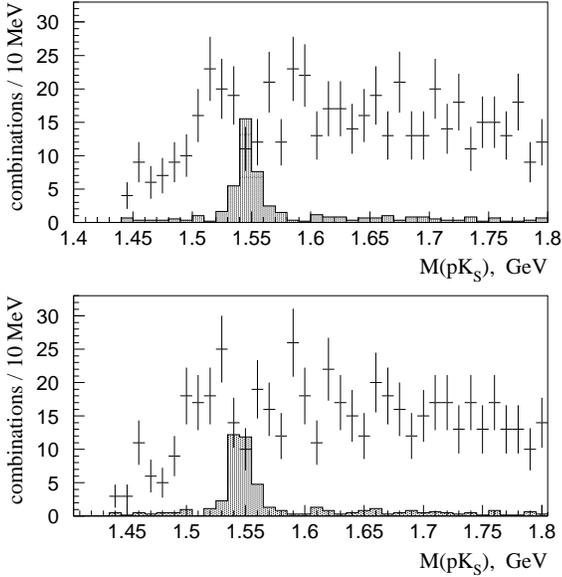}
\caption{%
Effective mass spectrum of $pK_S^0$ system for the reaction 
$p+N \to pK_S^0K_S^0 + N$. 
Lower picture corresponds to bin shift by 5~MeV. The MC
signal corresponds to the cross section ratio 
$\sigma_{\Theta^+\bar{K}^0}/\sigma_{\Lambda(1520)K^+}=$~0.1 
}
\label{fig-pksks}
\end{figure}

The results of the simulation of an interesting case of the $pK_S^0K_S^0$ 
final
state in comparison with the real data are shown in Fig.~\ref{fig-pksks-1}
and one-dimensional plots, used to set upper limits, in Fig.~\ref{fig-pksks}.

\begin{figure}
\includegraphics[width=\hsize]{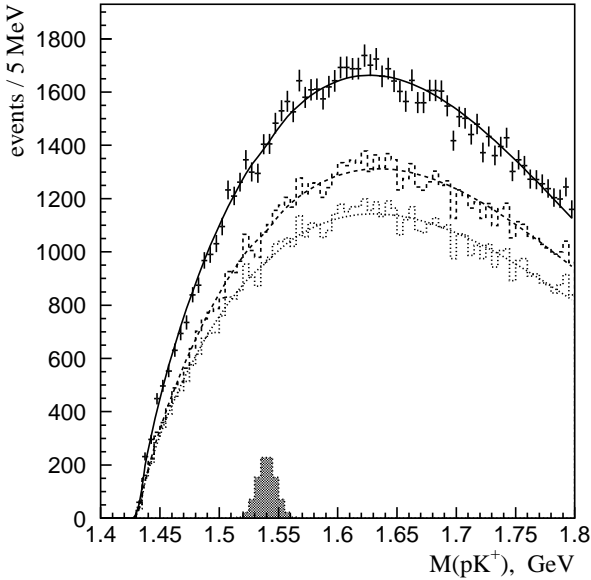}
\caption{%
Search for $\Theta^{++}$ baryon in the reaction 
$p + N \to \Theta^{++}K^- +N$; $\Theta^{++} \to pK^+$. The histogramms
are with no cuts, with $\Lambda(1520)$ and $\Lambda(1520)+\phi$ cuts.
The curves are results of the fit described in the test.
The MC signal from narrow $\Theta^{++}$ corresponds to the cross section ratio 
$\sigma_{\Theta^{++}K^-}/\sigma_{\Lambda(1520)K^+}=$~0.01 and 
$BR(\Theta^{++}\to pK^+) = 100\%$. 
}
\label{fig-pkplus}
\end{figure}

We also made a search for the $\Theta^{++}(1540)$ baryon in the
reaction $p + N \to \Theta^{++}K^- +N$; $\Theta^{++} \to pK^+$. 
This state emerges as a partner of $\Theta^+$ in some theoretical models.
The study was a byproduct of the investigation of the calibration reaction
$p+N \to pK^+K^- +N$. The $pK^+$ effective mass distributions are shown in
Fig.~\ref{fig-pkplus} for all events, for the events 
with ``$\Lambda(1520)$-cut''
(the events with $1.50 <M(pK^-)<1.55$~GeV were excluded), and events
with combined ``$\Lambda(1520)+\phi$-cut'', where in addition 
the events with $1.01 <M(K^+K^-)<1.03$~GeV were excluded. Neither
distribution shows statistically significant signal and only the
upper limits can be set for $\Theta^{++}K^-$ production.


To get the limit on the number of events from the distributions 
Figs.~\ref{fig-pkskl}, \ref{fig-nkplus}, \ref{fig-pksks} and \ref{fig-pkplus},
we tried two different methods. In the
first method the distributions were fitted by a smooth curve plus a gaussian
function with fixed mass and a resolution, determined by
MC simulation for a particular final state. The 90\% confidence level
upper limits 
for the number of events were then estimated in a usual way. 
In the second method the events in the mass
window near the presumed signal region were excluded from the fit and 
90\% confidence level upper limits were calculated as 
$N = \mathrm{max}(0,\Delta n) + 1.28\sqrt{b}$ with 
$\Delta n$ being the excess 
over estimate 
background $b$. The mass window was from 20~MeV for $[pK_S^0]K_L^0$ system 
to 40~MeV for $[pK_L^0]K_S^0$ one. To understand systematics, we 
varied the fitting range and background function. In the first method we
also made fits with MC resolution enlarged by 1~MeV. 
Both methods give similar results with no more than 20\% difference in
upper limits. In the rest of the paper we are using the results from 
the first method. 

\section{Results}
\label{sec-res}
Our results in a search for narrow exotic baryons are presented in 
Table~\ref{tab-3}, where efficiencies, gaussian
resolutions, fitted number of events and upper limits on 
cross section
for the reactions $p+N\to\Theta^+\bar{K}^0+N$ and $p+N\to\Theta^{++}K^-+N$ 
can be found.

The upper limits for production cross sections are preliminary. They
were calculated from the upper limits on the number of events using the
procedure described above for the calibration reactions. The measured
values were corrected with the same correction factor 
$k_{\mathrm{corr}} = 1.38$. The systematic errors for the production cross
sections are estimated to be $<25$\%.   
\begin{table*}[!t]
\begin{center}
\begin{tabular}{|cccccc|}
\hline\noalign{\smallskip}
 Particle       & Final state & Events & $\varepsilon$, &  Resolution & $\sigma_{\mathrm{corr}}$, 90\% CL \\ 
                &            &         & \%         &  ($\sigma$), MeV & nb/nucleon \\ 
\noalign{\smallskip}\hline\noalign{\smallskip}
$\Theta^+(1540)$& $[nK^+]K_S^0$ & $55\pm43$  & 3.3    &   9.8 & $<32$   \\
                &              & $10\pm33$  & 2.2   & 10.1 & $<26$  \\
                &              & ($\Lambda_{1520}$ cut)  &      &      & \\ \hline
                & $[pK_S^0]K_L^0$  & $48\pm29$  & 3.0   &  5.8 & $<53$   \\
                &              & $26\pm25$  & 2.7   &  5.5 & $<42$  \\
                &              & ($\phi$ cut) &     &      &        \\ \hline
                & $[pK_L^0]K_S^0$ & $6\pm43$  & 2.6     & 11.8 & $<54$   \\
                &              & $-14\pm37$  & 2.3  & 11.3 & $<39$  \\
                &              & ($\phi$ cut) &     &      &        \\ \hline
                & $[pK_S^0]K_S^0$  & $-4\pm7$  & 0.9    &  7.5 & $<52$   \\ \hline
$\Theta^{++}(1540)$ & $[pK^+]K^-$ & $-57\pm100$  & 10.0  &  8.0 & $<2$   \\ 
\noalign{\smallskip}\hline
\end{tabular}
\end{center}
\caption{%
Upper limits for reactions 
$p+N\to \Theta^+\bar{K^0}+N$ and $p+N\to \Theta^{++}K^- +N$ 
at $E_p=70$~GeV. The number of events corresponds to the fit of the 
distributions with smooth background function plus gaussian with 
$M = 1540~$MeV and MC resolution. The efficiencies was calculated
assuming $\mathrm{BR}(K_S^0\to\pi^+\pi^-) = 100\%$.
}
\label{tab-3}
\end{table*}

\begin{figure}
\includegraphics[width=\hsize]{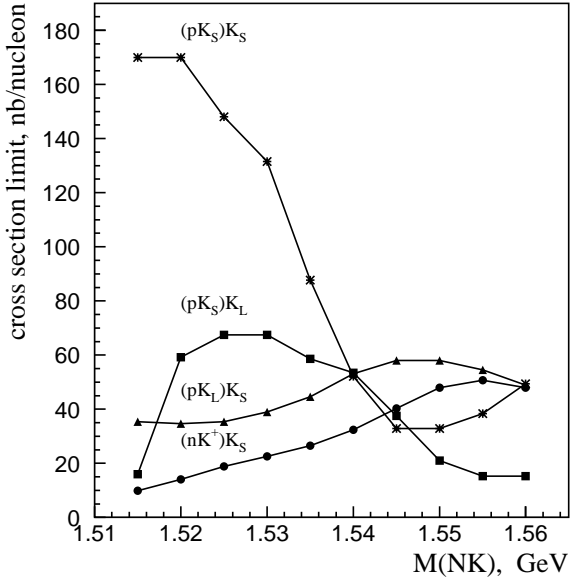}
\caption{%
Cross section limits on the $\Theta^+$ baryon production for
different masses of $\Theta^+$.
}
\label{fig-limits}
\end{figure}

Since different ``positive'' experiments reported different
masses for the $\Theta^+$ baryon, we made a scan in the effective mass for 
each 
of the different final states. The results of the scan are presented in 
Fig.~\ref{fig-limits}. A poor upper limit for the $[pK_S^0]K_S^0$ final state
in the mass range 1515-1530~MeV is a consequence of a bump in $pK_S^0$
mass spectrum, which is partly supported by a shoulder in $pK_S^0$ mass
spectrum of $[pK_S^0]K_L^0$ final state. However, an interpretation of this
bump as a (relatively wide) $\Theta^+$ baryon is absolutely excluded by $nK^+$
data. It is also possible that one-star $\Sigma(1480)$-hyperon really
exists, producing the irregularities in $pK_S^0$ mass spectrum (but why 
not in $pK_L^0$?).

We tried to apply additional cuts in order to find a signal.
These cuts included, in particular:
\begin{itemize}
\item
Cut on transverse momentum of the $NK\bar{K}$ system 
(coherent region $P_T^2<0.1~\rm{GeV}^2/c^2$, high $P_T^2$ region, etc).
\item
Cut on transverse momentum of the $NK$ system. 
\item
Cut on the effective mass of the $NK\bar{K}$ system.
\item
Cuts on the decay angles $(\cos\theta^*,\phi^*)$ of
the $[NK]\bar{K}$ system.
\end{itemize}
None of these cuts allowed us to see a statistically significant
signal in the region of the $\Theta^+$ baryon. 
In addition we don't see any evidence for the existence of narrow
structures in $NK\bar{K}$ effective mass distributions, in particular 
the state with $M\approx2.4$~GeV, indicated by CLAS 
data~\cite{Kubarovsky:2003fi}. It should be
noted, that with the extreme values of cuts, when statistics is started
to be low, irregularities in different mass spectra begin to appear,
sometimes at a mass of the $\Theta^+$ baryon. In our case, however, they 
never occured
simultaneously in the same place under the same cuts in more then one 
mass spectrum.

\begin{figure}
\includegraphics[width=\hsize]{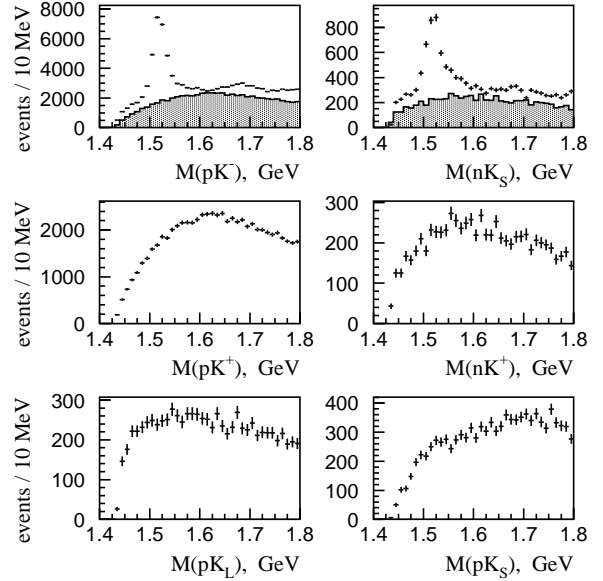}
\caption{%
A summary plot with $M(NK/\bar{K})$-distributions for $S=-1$ (upper row),
$S=+1$ (middle row) and mixt $S$ (lower row). The distributions for $S=+1$
are superimposed on the distributions for $S=-1$ as hatched histograms.
}
\label{fig-summary}
\end{figure}

Our distributions 
for ``allowed'' ($S=-1$),
``exotic'' ($S=+1$) and ``mixt'' (unknown) strangeness are compiled
in Fig.~\ref{fig-summary}. The compilation includes $pK^-$, $nK_S^0$,
$pK^+$, $nK^+$, $pK_L^0$ and  $pK_S^0$ mass spectra from 
$pK^+K^-$, $nK_S^0K^+$ and $pK_S^0K_L^0$ final states. 
In the distributions for ``exotic'' and ``mixt'' strangeness   
events, corresponding to the production of $\phi$, $\Lambda(1520)$, or both,
were excluded.


The relative yield $\Theta^+/\Lambda(1520)$ is the most commonly 
used variable to compare different experiments. In our case it transforms
to the cross section 
ratio $\sigma(\Theta^+\bar{K}^0)/\sigma(\Lambda(1520)K^+)$. The best way
to calculate this ratio is to use the $nK_S^0K^+$ final state, which is common
to both systems. Many factors cancel in the ratio and we
arrive to ($\Lambda^*\equiv \Lambda(1520)$)
\begin{eqnarray}
\label{eq_nkplus_ratio}
R_{\Lambda^*} \equiv 
\frac{\sigma(\Theta^+\bar{K}^0)}{\sigma(\Lambda^*K^+)} = & & \nonumber \\ 
\hspace*{-5mm}=\frac{N_{\Theta}}{N_{\Lambda^*}}\cdot 
\frac{\mathrm{BR}(\Lambda^*\to N\bar{K})}{\mathrm{BR}(\Theta^+\to NK)}\cdot 
\frac{\varepsilon_{\Lambda^*}}{\varepsilon_{\Theta^+}} = 
0.45\cdot \frac{N_{\Theta}}{N_{\Lambda^*}}\cdot
\frac{\varepsilon_{\Lambda^*}}{\varepsilon_{\Theta^+}}, & &
\end{eqnarray}
where we used PDG value $\mathrm{BR}(\Lambda^*\to N\bar{K})$ =~45\% and
$\mathrm{BR}(\Theta^+\to NK)$ was assumed to be 100\%. Number
of events and efficiencies for corresponding decay modes can be read
from Tables~\ref{tab-2} and \ref{tab-3}, and we finally get
\begin{equation}
\label{eq_nkplus_ratio_2}
R_{\Lambda^*} = 
0.45\cdot \frac{55\pm 43}{2490\pm 90}\cdot
\frac{0.038}{0.033} = 0.011\pm 0.007, \nonumber
\end{equation}
with an upper limit  $R_{\Lambda^*}<$~0.02~ at 90\% CL for 
$M(\Theta^+) = 1540$~MeV. The mass dependence of this limit
can be understood from $[nK^+]K_S^0$ curve in Fig.~\ref{fig-limits}.
This estimate is free of any correction factors, and it is hard to imagine
that the efficiency of $\Theta^+$ detection is significantly lower than 
that of $\Lambda(1520)$. Note that for 
$\sigma(\Theta^+\bar{K}^0)\approx\sigma(\Lambda(1520)K^+)$ and 
$\varepsilon_{\Theta^+}\approx\varepsilon_{\Lambda^*}$ we should see 
as much as 5000 $\Theta^+\to nK^+$ decays. 

The ratio  $R_{\phi} \equiv 
\sigma(\Theta^+\bar{K}^0)/\sigma(p\phi)$ can be estimated in the 
same way as $R_{\Lambda^*}$ and the result is
$R_{\phi}<$~0.15 at 90\% CL.
\section{Discussion}

Our result for $R_{\Lambda^*}$ is based on the sample of 
${\approx}2500~\Lambda(1520)\to nK_S^0$ decays or, equivalently, on
${\approx}21000~\Lambda(1520)\to pK^-$ decays, corresponding to
$\approx 900$k of $\Lambda(1520)$ produced in the reaction
$p+N\to \Lambda(1520)K^++N$.
With somewhat smaller number of  $\Lambda(1520)$
HERA-B~\cite{HERA-B} reports (preliminary) $R_{\Lambda^*}<$~0.02
for inclusive $pA$ interactions
at 920~GeV/$c$ and mid-rapidity (${\approx}6000~\Lambda(1520)\to pK^-$ decays)
and 
ALEPH~\cite{ALEPH} gives $R_{\Lambda^*}<$~0.1 for inclusive decays of $Z$,
with a sample of 2819 $\Lambda(1520)\to pK^-$ decays.
 
Low values of $R_{\Lambda^*}$ in these ``negative'' experiments are 
in a striking contrast with this ratio (when available) for ``positive'' 
ones. In  quasi-real photoproduction $\gamma^*d\to \Theta^+X$
HERMES~\cite{Airapetian:2003ri} found $R_{\Lambda^*}=$~1.6--3.5 with
$\approx 800~\Lambda(1520)\to pK^-$ events and SAPHIR~\cite{Barth:2003es}
in exclusive reaction $\gamma p \to nK^+K_S^0$ estimated 
$\sigma(\Theta^+K_S^0)\approx$~300~nb, corresponding to 
$R_{\Lambda^*}\approx$~1/3.

A small value of $R_{\Lambda^*}$, found by HERA-B and SPHINX in 
proton-induced reactions, corresponds, however, to quite different 
physical processes and
kinematical regimes. It is natural to assume that this value
should hold for any proton(nucleon)-induced reactions. 
It would be interesting to have this ratio for the ``positive''
experiments for nucleon-induced reactions.  
Unfortunately, $\Lambda(1520)$ production was not studied by 
SVD-2 collaboration~\cite{Aleev:2004sa}, where 
some signal for $\Theta^+\to pK_S^0$ was found in 
inclusive $pA$ interactions at 70 GeV/$c$. The full cross section 
of $\Theta^+$
production in the fragmentation region ($X_F(pK_S^0)>$~0) was found to be
(30--120) $\mu$b/nucleon. We can compare this rather big cross section
with the measurements of the inclusive production of $\Lambda(1520)$
in neutron/nucleus interactions at 40~GeV/$c$~\cite{Krastev:1988xr},
where the cross section of $\Lambda(1520)$ production was found to
be (70--90)~$\mu$b/nucleon for $X_F(\Lambda(1520))>$~0. Assuming the
ratio for the inclusive cross sections of $\Lambda(1520)$ production 
for the beam momenta 70 and 40~GeV/$c$ to be 1.5 (for
well studied $\Lambda$ production the corresponding factor is 
${\approx}1.4$), we conservatively find 
$R_{\Lambda^*} >~30/(90\cdot 1.5) = 0.22$.

In the other experiment~\cite{Aslanyan:2004gs}, a positive signal was found
in the $pK_S^0$ system in $p+\mathrm{C}_3\mathrm{H}_8$ collisions 
at 10 GeV/$c$. Preliminary
cross section for $\mathrm{C}_3\mathrm{H}_8$ molecule was found 
to be 90~$\mu$b, corresponding
to $\approx$7--9~$\mu$b/nucleon (our estimate). The production of 
$\Lambda(1520)$ was not reported. We can estimate $R_{\Lambda^*}$ 
for this experiment using
the results of~\cite{Ansorge:1974ez}, where the inclusive
cross section for $\Lambda(1520)$ production was measured in $n\,p$ 
interactions at 4--8~GeV/$c$ and found to be $\approx 7\,\mu$b/nucleon. 
A conservative estimate gives $R_{\Lambda^*}>$~0.2--0.3.    

Thus the results~\cite{Aleev:2004sa,Aslanyan:2004gs}  seem to be in 
direct contradiction with HERA-B and SPHINX data.   

Our upper limit on the absolute cross section production of $\Theta^+$
can be used to set upper limit on its width, though in a model
dependent way. The forward hemisphere cross section for
the reaction $pp\to \Theta^+\bar{K}^0p$ was calculated 
in~\cite{Liu:2003rh} using hadronic lagrangian with empirical coupling
constants and form factors. The cross section was found to have a maximum
of 13~$\mu$b at $\sqrt{s}\approx$~4.5~GeV. The last point in their figure
at $\sqrt{s}=$~7~GeV gives the value $\approx$6.5~$\mu$b. We can extrapolate
this value for $\sigma(\Theta^+\bar{K}^0)$ to our energy 
($\sqrt{s}=$~11.5~GeV) using (conservatively) $1/P_{\mathrm{beam}}^2$
dependence and find  $\sigma(\Theta^+\bar{K}^0)\approx 900$~nb.
As both cross section and total width of $\Theta^+$ are proportional
to $g^2_{\Theta NK}$ and the width of $\Theta^+$ was assumed to be
$\Gamma[\Theta^+]=$20~MeV in the calculations, our upper limit on 
the production
cross section transforms to the upper limit on the width, 
$\Gamma[\Theta^+]<20\cdot(32/900) = 0.7$~MeV (the limit from $nK^+$
decay mode was used). This result is in a good 
agreement with constraints on $\Gamma[\Theta^+]$ deduced from
$K^+N$ and $K^+d$ data~\cite{ThetaWidth}. It should be noted, in 
addition, that with $\Gamma[\Theta^+]<0.7$~MeV the 
calculations~\cite{Liu:2003rh} would give the cross section
for the reaction $pp\to \Theta^+\Sigma^+<$ few tens of nb, 
in a disagreement with COSY-TOF~\cite{Abdel-Bary:2004ts} measurements
of $400\pm100\pm100$~nb/nucleon for this reaction at P=2.95 GeV/$c$.


\section{Conclusion}
In a high statistics experiment with the SPHINX 
facility at IHEP accelerator
we have searched for the production
of the $\Theta^+$ baryon in exclusive reaction 
$p+N\to\Theta^+\bar{K}^0+N$
at the energy of 70~GeV. For the first time the search was done
simultaneously in all possible decay modes of $\Theta^+$.
We did not see statistically significant signals and found 
that the production of the $\Theta^+\bar{K}^0$ system  
is very small (if any) compared to the production of 
$\Lambda(1520)K^+$ system and small (if any) compared to OZI-suppressed
production of the $p\phi$ system.

The work is underway in the search for $\Theta^+$ in the reaction
$p+N \to \Theta^+ K^-\pi^++N$\footnote
{When this work was already written, we find out that E690 
collaboration at Fermilab have searched for the $\Theta^+$ baryon in the
similar reaction 
$pp\to [p_{\mathrm{slow}}K_S^0]K^-\pi^+p_{\mathrm{fast}}$ at 
800~GeV/c, with a negative result. Their preliminary results can be
found in~\cite{E690}.
}.
In a search for the exotic mechanisms for $\Theta^+$ production
we also plan to study the final states with baryon-antibaryon pair,
like 
$p + N \to \Theta^+ \Sigma^+\bar{p} + N$. 
\begin{acknowledgement}
It is a pleasure to express our gratitude to the  staff of IHEP 
accelerator and especially A.A.~Aseev for the invaluable 
help with the beam adjustment in the critical moments of data taking, and
to A.M.~Zaitsev for useful discussions. 
This work was partly supported by Russian
Foundation for Basic Researches 
(grants 99-02-18252 and~02-02-16086).
\end{acknowledgement}


\begin{thebibliography}{}

\bibitem{Nakano:2003qx}
T.~Nakano {\it et al.}  [LEPS Coll.],
Phys.\ Rev.\ Lett.\  {\bf 91}, 012002 (2003), hep-ex/0301020.
\bibitem{Barmin:2003vv}
V.~V.~Barmin {\it et al.}  [DIANA Coll.],
Phys.\ Atom.\ Nucl.\  {\bf 66}, 1715 (2003)
[Yad.\ Fiz.\  {\bf 66}, 1763 (2003)], hep-ex/0304040.
\bibitem{Stepanyan:2003qr}
S.~Stepanyan {\it et al.}  [CLAS Coll.],
Phys.\ Rev.\ Lett.\  {\bf 91}, 252001 (2003), hep-ex/0307018.
\bibitem{Barth:2003es}
J.~Barth {\it et al.}  [SAPHIR Coll.],
hep-ex/0307083.
\bibitem{Asratyan:2003cb}
A.~E.~Asratyan, A.~G.~Dolgolenko and M.~A.~Kubantsev,
hep-ex/0309042.
\bibitem{Kubarovsky:2003fi}
V.~Kubarovsky {\it et al.}  [CLAS Coll.],
[Phys.\ Rev.\ Lett.\  {\bf 92}, 032001 (2004)]
Erratum -- ibid.\  {\bf 92}, 049902 (2004),
hep-ex/0311046.
\bibitem{Togoo}
R.~Togoo {\it et al.}, Proc. Mongolian Acad. Sci., {\bf 4} (2003) 2.
\bibitem{Airapetian:2003ri}
A.~Airapetian {\it et al.}  [HERMES Coll.],
Phys.\ Lett.\ B {\bf 585}, 213 (2004)
hep-ex/0312044.
\bibitem{Aleev:2004sa}
A.~Aleev {\it et al.}  [SVD Coll.],
hep-ex/0401024.
\bibitem{Abdel-Bary:2004ts}
M.~Abdel-Bary {\it et al.}  [COSY-TOF Coll.],
hep-ex/0403011.
\bibitem{Aslanyan:2004gs}
P.~Z.~Aslanyan, V.~N.~Emelyanenko and G.~G.~Rikhkvitzkaya,
hep-ex/0403044.
\bibitem{Chekanov:2004kn}
S.~Chekanov {\it et al.}  [ZEUS Coll.],
hep-ex/0403051.
\bibitem{Dzierba:2003cm}
A.~R.~Dzierba, D.~Krop, M.~Swat, S.~Teige and A.~P.~Szczepaniak,
Phys.\ Rev.\ D {\bf 69} (2004) 051901,
hep-ph/0311125.
\bibitem{Rosner:2003ia}
J.~L.~Rosner,
Phys.\ Rev.\ D {\bf 69} (2004) 094014,
hep-ph/0312269.
\bibitem{Zavertyaev:2003wv}
M.~Zavertyaev,
hep-ph/0311250.
\bibitem{Zhao:2004rg}
Q.~Zhao and F.~E.~Close,
hep-ph/0404075.
\bibitem{Klempt:2004yz}
E.~Klempt,
hep-ph/0404270.

\bibitem{BES}
J.~Z.~Bai {\it et al.}  [BES Collaboration],
hep-ex/0402012.
\bibitem{HERA-B}
K.~T.~Knopfle, M.~Zavertyaev and T.~Zivko  [HERA-B Coll.],
hep-ex/0403020.
\bibitem{PHENIX}
C.~Pinkenburg [for the PHENIX Coll.],
nucl-ex/0404001.
\bibitem{ALEPH}
P. Hansen [for ALEPH Coll.], talk at DIS 2004,
\hfill\break
{\tt http://www.saske.sk/dis04/talks/C/hansen.pdf}
\bibitem{DELPHI}
Throsten Wengler [reporting DELPHI Coll. results], talk at
Moriond '04 QCD,
http://moriond.in2p3.fr/QCD/2004/
WednesdayAfternoon/Wengler.pdf.
\bibitem{Karliner:2004gr}
M.~Karliner and H.~J.~Lipkin,
hep-ph/0405002.
\bibitem{Balats:zf}
M.~Y.~Balats {\it et al.}  [SPHINX Collaboration.],
Z.\ Phys.\ C {\bf 61} (1994) 223.
V.~A.~Dorofeev {\it et al.}  [SPHINX Collaboration],
Phys.\ Atom.\ Nucl.\  {\bf 57} (1994) 227
[Yad.\ Fiz.\  {\bf 57} (1994) 241].
\bibitem{Vavilov:1993cv}
D.~V.~Vavilov {\it et al.}  [SPHINX Collaboration.],
Phys.\ Atom.\ Nucl.\  {\bf 57} (1994) 1970
[Yad.\ Fiz.\  {\bf 57} (1994) 2046].
\bibitem{Balats:xh}
M.~Y.~Balats {\it et al.}  [SPHINX Collaboration.],
Z.\ Phys.\ C {\bf 61} (1994) 399.
\bibitem{Golovkin:1999ww}
S.~V.~Golovkin {\it et al.}  [SPHINX Collaboration],
Eur.\ Phys.\ J.\ A {\bf 5} (1999) 409.
\bibitem{Vavilov:2000vw}
D.~V.~Vavilov {\it et al.}  [SPHINX Collaboration],
Phys.\ Atom.\ Nucl.\  {\bf 63} (2000) 1391
[Yad.\ Fiz.\  {\bf 63} (2000) 1469].
\bibitem{Landsberg:1999gr}
L.~G.~Landsberg,
Phys.\ Rept.\  {\bf 320} (1999) 223.
\bibitem{Antipov:eh}
Y.~M.~Antipov {\it et al.}  [SPHINX Collaboration],
Phys.\ Atom.\ Nucl.\  {\bf 65} (2002) 2070
[Yad.\ Fiz.\  {\bf 65} (2002) 2131].
\bibitem{Diakonov:1997mm}
D.~Diakonov, V.~Petrov and M.~V.~Polyakov,
Z.\ Phys.\ A {\bf 359} (1997) 305,
hep-ph/9703373.
\bibitem{Weigel:1998vt}
H.~Weigel,
Eur.\ Phys.\ J.\ A {\bf 2} (1998) 391,
hep-ph/9804260.
\bibitem{Antipov:1994fn}
Y.~Antipov {\it et al.},
Nucl.\ Phys.\ Proc.\ Suppl.\  {\bf 44} (1995) 206.
\bibitem{Kozhevnikov:rv}
A.~Kozhevnikov, V.~Kubarovsky, V.~Molchanov, V.~Rykalin and V.~Solyanik,
Nucl.\ Instrum.\ Meth.\ A {\bf 433} (1999) 164.
\bibitem{Powell:1981it}
B.~Powell {\it et al.}  [CERN-Heidelberg-Padua-Paris-Rome-Serpukhov-Trieste
                  Collaboration],
Nucl.\ Instrum.\ Meth.\  {\bf 198}, 217 (1982).
\bibitem{Antipov:1989vm}
Y.~M.~Antipov {\it et al.},
Nucl.\ Instrum.\ Meth.\ A {\bf 295}, 81 (1990).
\bibitem{Bityukov:1994ij}
S.~I.~Bityukov {\it et al.},
IFVE-94-101
%
\bibitem{Hagiwara:fs}
K.~Hagiwara {\it et al.}  [Particle Data Group Collaboration],
Phys.\ Rev.\ D {\bf 66} (2002) 010001.
\bibitem{Arenton:ts}
M.~W.~Arenton, D.~S.~Ayres, R.~Diebold, E.~N.~May, L.~Nodulman, J.~R.~Sauer and A.~B.~Wicklund,
Phys.\ Rev.\ D {\bf 25} (1982) 22.
\bibitem{Golovkin:1997uu}
S.~V.~Golovkin {\it et al.}, [SPHINX Collaboration.],
Z.\ Phys.\ A {\bf 359} (1997) 435.
\bibitem{Krastev:1988xr}
V.~R.~Krastev {\it et al.},
JINR-P1-88-31
\bibitem{Ansorge:1974ez}
R.~E.~Ansorge, J.~R.~Carter, J.~A.~Charlesworth, W.~W.~Neale and J.~G.~Rushbrooke,
Phys.\ Rev.\ D {\bf 10} (1974) 32.
\bibitem{Liu:2003rh}
W.~Liu and C.~M.~Ko,
Phys.\ Rev.\ C {\bf 68} (2003) 045203,
nucl-th/0308034.
\bibitem{ThetaWidth}
S. Nussinov,  hep-ph/0307357;
R.~W.~Gothe and S.~Nussinov,
hep-ph/0308230;
R.~A.~Arndt, I.~I.~Strakovsky and R.~L.~Workman,
Phys.\ Rev.\ C {\bf 68} (2003) 042201,
nucl-th/0308012 and nucl-th/0311030;
%
J. Haidenbauer and G. Krein,  hep-ph/0309243;
%
R.~N.~Cahn and G.~H.~Trilling,
hep-ph/0311245;
%
A.~Casher and S.~Nussinov,
Phys.\ Lett.\ B {\bf 578} (2004) 124,
hep-ph/0309208.
A.~Sibirtsev, J.~Haidenbauer, S.~Krewald and U.~G.~Meissner,
hep-ph/0405099.

\bibitem{E690}
D. Christian, E690 Collaboration, Quarks and Nuclear Physics 2004,
Bloomington, Indiana, May 23-28, 2004, http://www.qnp2004.org/

\end{thebibliography}
\end{document}